\documentstyle[12pt,ssi]{article}
\begin{document}
\renewcommand\floatpagefraction{1}
\renewcommand\topfraction{1}
\renewcommand\bottomfraction{1}
\renewcommand\floatsep{12pt}
\renewcommand\textfraction{0}
\renewcommand\intextsep{12pt}
\def\simpropto{\mathrel{\raise.3ex\hbox{$\propto$}\mkern-14mu
              \lower0.6ex\hbox{$\sim$}}}
\def\gtorder{\mathrel{\raise.3ex\hbox{$>$}\mkern-14mu
             \lower0.6ex\hbox{$\sim$}}}
\def\ltorder{\mathrel{\raise.3ex\hbox{$<$}\mkern-14mu
             \lower0.6ex\hbox{$\sim$}}}
\widowpenalty=20000
\clubpenalty=20000
\input psfig

\title{
An Introduction to Solar Neutrino Research\footnotemark}
\footnotetext{*These
lectures were presented at the XXV SLAC Summer Institute on Particle
Physics, ``Physics of Leptons,'' August 4--15, 1997. To be published
in a SLAC Report on the proceedings.}
\author{John Bahcall\\
Institute for Advanced Study, Princeton, NJ 08540}
\maketitle

\addvspace{1cm}
\centerline{\large PROLOGUE}
\bigskip
In the first lecture, I describe the conflicts between the combined
standard model 
predictions and the results of solar neutrino experiments.
Here `combined standard model' means the  minimal standard
electroweak model plus a standard solar model.
First, I show how  the comparison between standard model
predictions and the observed rates in the four pioneering experiments
leads to three different solar neutrino problems.
Next, I summarize the stunning
agreement between the predictions of standard solar models and 
helioseismological measurements; this precise agreement suggests that 
future refinements of solar model physics are unlikely to affect
significantly the three solar neutrino problems.
Then, I describe the important  recent analyses
in which the neutrino fluxes are treated as free parameters,
independent of any constraints from solar models.
The disagreement that exists even without using any solar model
constraints further reinforces the view that new physics may be required.
The principal conclusion of the first lecture is that the minimal standard
model is not consistent with the experimental 
results that have been reported for the pioneering solar
neutrino experiments.

In the second lecture, I discuss the possibilities for detecting
``smoking gun'' indications of departures from minimal standard electroweak
theory.  Examples of smoking guns are the 
distortion of the energy spectrum of recoil electrons produced by
neutrino interactions, the 
dependence of the observed
counting rate on the zenith angle of the sun (or, equivalently, the 
path through the earth to the detector), the ratio of the flux of
neutrinos of all types to the flux of electron  neutrinos (neutral
current to charged current ratio), and seasonal variations of the
event rates (dependence upon the earth-sun distance).

\section{Introduction}
\label{intro}
Solar neutrino research entered a new era in April, 1996, when the 
Super-Kamiokande experiment\cite{Takita93,Totsuka96} began to
operate.  
We are now in a period of precision, high-statistics tests of standard
electroweak theory and of stellar evolution models.

In the previous era, solar neutrinos were 
detected by four beautiful experiments, the radiochemical Homestake chlorine
experiment,\cite{Davis94,cleveland98} 
the Kamiokande water Cerenkov experiment,\cite{Suzuki95,kamioka} 
and the two
radiochemical gallium experiments, GALLEX\cite{Ansel95} and
SAGE.\cite{Abdur94} 
In these four exploratory experiments, typically less than or of the
order of 50 neutrino events were observed per year.

The pioneering 
 experiments achieved the scientific goal which was
set in the early 1960s,\cite{bahcall64,Davis64} 
namely, ``...to see into the interior of a
star and thus verify directly the hypothesis of nuclear energy
generation in stars.''  We now know from experimental measurements,
not just theoretical calculations, that the sun shines by nuclear
fusion among light elements, burning hydrogen into helium.

Large electronic detectors will yield vast amounts of diagnostic data
in the new era that has just begun.
Each of the new electronic experiments is expected to produce of order several
thousand neutrino events per year.
These experiments, Super-Kamiokande,\cite{Takita93,Totsuka96}
SNO,\cite{McD94} and BOREXINO\cite{Arp92}
will test the prediction of the minimal standard
electroweak model\cite{Glashow61,Wein67,Salam68} 
that essentially nothing happens to electron 
neutrinos after they are created by nuclear fusion reactions in the
interior of the sun.

The four pioneering experiments---chlorine\cite{Davis94,cleveland98,Davis64} 
Kamiokande\cite{Suzuki95,kamioka} 
GALLEX\cite{Ansel95}  and SAGE\cite{Abdur94}---have 
all observed neutrino fluxes with intensities that are within a
factors of a few of  
those predicted by standard solar models. 
Three of the experiments (chlorine, GALLEX, and SAGE) are
radiochemical and each radiochemical experiment  measures
 one number, the total rate at which
neutrinos above a fixed energy threshold (which depends upon the
detector)  are detected.  
The sole electronic (non-radiochemical) 
detector among the initial experiments,
Kamiokande, has shown 
that the neutrinos come from the sun,
by measuring the recoil directions of the  electrons scattered by
solar neutrinos.
Kamiokande has also demonstrated 
that the observed neutrino energies 
are consistent with 
 the range of energies expected on the basis of the standard solar model.

Despite continual refinement of solar model calculations of
neutrino fluxes over the past 35 years (see, e.g., the collection of
 articles reprinted in the book edited by 
Bahcall, Davis, Parker, Smirnov, and Ulrich\cite{BDP95}),
the discrepancies between 
observations and calculations have gotten worse with time.  All four
of the pioneering solar neutrino experiments yield event rates that
are significantly less than predicted by standard solar models.

These lectures are organized as follows.
I first discuss in Section~\ref{threeproblems} the three solar
neutrino problems. Next I discuss in Section~\ref{helioseismology} 
the stunning agreement
between the values of the sound speed calculated from standard
solar models and the values obtained from helioseismological
measurements. 
Then I review 
in Section~\ref{lasthope}
 recent work which 
treats the neutrino fluxes as free parameters and 
shows that the solar neutrino problems 
cannot be resolved within the context of the minimal standard electroweak
model unless some solar neutrino experiments are incorrect.
At this point, I summarize in Section~\ref{closing} the main conclusions of
the first lecture. I begin the second lecture by describing in
Section~\ref{newexperiments} 
the new solar neutrino experiments and then answer in
Section~\ref{whyphysicists}
the question: Why do physicists care about solar
neutrinos?
I present briefly in Section~\ref{allowedmsw} and Section~\ref{vacuum},
respectively, 
the MSW solutions and the vacuum oscillation solutions 
that describe well the
results of the four pioneering solar neutrino experiments.
Finally, in Section~\ref{smoking} I describe
the ``smoking gun'' signatures of physics beyond the minimal standard
electroweak model that are being searched for with the new solar
neutrino detectors. I summarize in Section~\ref{summarytwo} 
my view of where we are now in solar
neutrino research.

I will concentrate in Lecture~I on comparing  the predictions of the combined
standard model with the results of the  operating solar neutrino
experiments.    
By `combined' standard model, I mean the predictions of the standard
solar model and the predictions of the minimal standard electroweak theory.

We need a solar model to tell us how many neutrinos of what energy 
are produced per unit of time in the sun.
Our physical intuition is not yet sufficiently
advanced to know if we should be surprised by $10^{-2}$, by $10^{0}$,
or by $10^{+2}$
neutrino-induced  events per day in a chlorine tank the size of an
Olympic swimming pool.
Specifically, solar model calculations are  required in order to predict
 the rate of nuclear fusion by the \hbox{$pp$} chain (shown in
Table~\ref{ppreactions} 
and the rate of fusion by the CNO reactions (originally favored by H.
Bethe in his epochal study of nuclear fusion reactions).
In a modern standard solar model, about $99$\% of the energy generation is
produced by reactions in the $pp$ chain. The
most important neutrino producing reactions 
(cf. Table~\ref{ppreactions}) are the 
low energy $pp$, $pep$, and 
the $^7$Be  neutrinos, and the higher energy $^8$B neutrinos.

\begin{table}[htb]
\centering
\caption[]{The Principal Reactions of the $pp$ Chain}
\begin{tabular}[htb]{lcc}
\noalign{\medskip}
\hline\hline
\noalign{\smallskip}
\multicolumn{1}{c}{Reaction}&Reaction&Neutrino Energy\\ 
Number&&(MeV)\\ 
\noalign{\smallskip}
\hline
\noalign{\medskip}
1&$\phantom{^3}p + p \to {\rm ^2H} + e^+ + \nu_{e}$&0.0 to 0.4 \\ 
2&\ \ \ \ $\phantom{^8}p + e^- + p \to {\rm ^2H} + \nu_{e}$&1.4 \\ 
3&${\rm ^2H} + p \to {\rm ^3He} + \gamma$& \\ 
4&${\rm ^3He} + {\rm ^3He} \to {\rm ^4He} + 2p$& \\ 
&\multicolumn{1}{c}{or} \\
5&${\rm ^3He} + {\rm ^4He} \to {\rm ^7Be} + \gamma$& \\ 
&\multicolumn{1}{c}{then} \\ 
6&\ \ \ \ $\phantom{^3}e^- + {\rm ^7Be} \to {\rm ^7Li} + \nu_{e}$&0.86,
0.38 \\ 
7&\ \ \ \ \ \ \ \ ${\rm ^7Li} + p \to {\rm ^4He} + {\rm ^4He}$& \\ 
&\multicolumn{1}{c}{or} \\ 
8&\ \ \ \ $\phantom{^3}p + {\rm ^7Be} \to {\rm ^8B} + \gamma$& \\ 
9&\ \ \ \ \ \ \ \ ${\rm ^8B} \to {\rm ^8Be} + e^+ + \nu_{e}$&0 to 15 \\ 
\noalign{\medskip}
\hline
\end{tabular}
\smallskip
\label{ppreactions}
\end{table}

A  particle
physics model is required to predict what happens to the neutrinos after
they are created, whether or not  
flavor content of the neutrinos is changed as they
make their way from the center of the sun to detectors on earth.
For the first part of our discussion, I assume
 that essentially nothing happens to the neutrinos
after they are created. In particular, they do not oscillate or decay
to neutrinos with a different lepton number or energy.
 This assumption is valid if minimal
standard electroweak theory is correct.  In the simplest version of
standard electroweak theory, neutrinos are massless and
neutrino flavors (the number of $\nu_e$ or $\nu_mu$ or $\nu_tau$) are
separately  conserved.  The minimal
standard electroweak model has had many successes in precision 
 laboratory tests;  modifications of this theory will be
accepted only if 
incontrovertible experimental evidence forces a change.

We will see that this comparison between combined standard model and
solar neutrino experiments leads to three
different discrepancies between the calculations and the observations,
which I will refer to as the three solar neutrino problems.
In the next section, I will discuss each of these three problems.

This is not a review article.  My goal is 
to describe where we stand in solar neutrino research and where
we are going, not to systematically describe the published literature.
Some of the relevant background is presented in the excellent lectures in this
Summer School by M. Davier, H. Harari, K. Martens, and S. Wojciki.
See my home page
http://www.sns.ias.edu/\raise2pt\hbox{${\scriptstyle\sim}$}jnb 
for more complete information about solar neutrinos, including 
anotated viewgraphs,
preprints, and numerical data.
Additional introductory material at roughly the level presented here
can be found in two other recently published lectures.\cite{bahcall96,bahcall97}
I have used here some material from these earlier talks but unfortunately
could not cover everything contained in the previous discussions.

\section{Three Solar Neutrino Problems}
\label{threeproblems}

\begin{figure}[t]
\hglue.5in{\psfig{figure=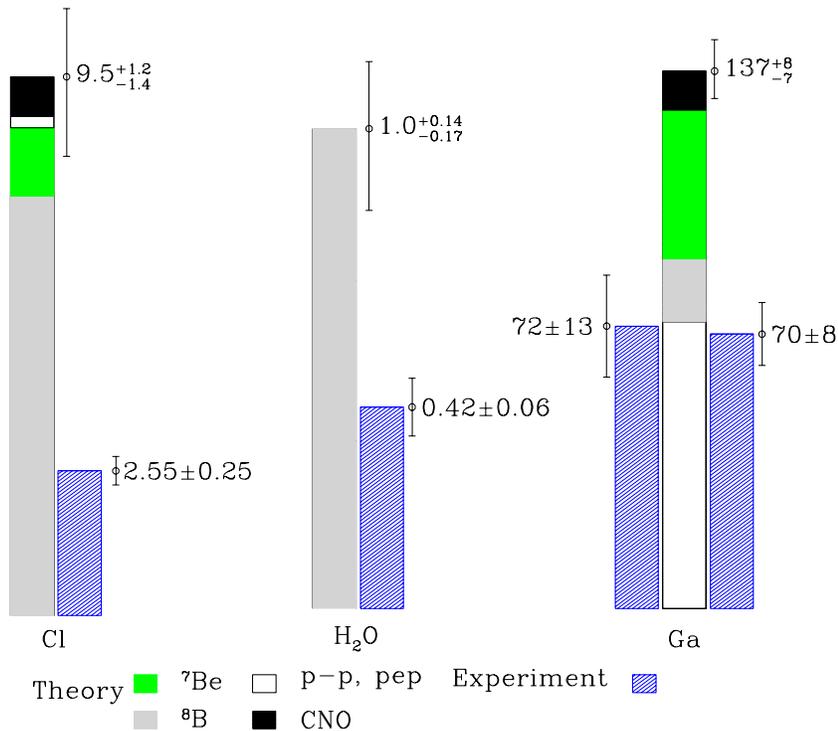,width=4.5in}}
\caption[]{\baselineskip=10pt{\footnotesize Comparison of measured 
rates and 
standard-model predictions
for four solar neutrino experiments\label{compare}.}\smallskip}
\end{figure}

Figure~\ref{compare} shows almost everything currently known about
the solar solar neutrino problems.

The figure compares the measured and the calculated event
rates in the four pioneering experiments,  
revealing three discrepancies between the 
experimental results and the expectations based
upon the combined standard model.  As we shall see, only
the first of these discrepancies depends sensitively upon
predictions of the standard solar model.

\subsection{Problem 1. Calculated versus Observed Absolute Rate}
\label{firstproblem}

The first solar neutrino experiment to be performed was the chlorine
radiochemical experiment, which detects electron neutrinos that
are  more energetic 
than $0.81$ MeV.  After more than
25 years of the operation of this experiment,\cite{cleveland98} 
the measured event
rate is $2.55 \pm 0.25$ SNU\footnote{If you appreciate
experimental beauty, courage, and ingenuity, 
then you must read the epochal paper by Cleveland,
Davis, Lande, and their collaborators in which they describe three
decades of ever more precise measurements with the Homestake chlorine
neutrino experiment.\cite{cleveland98}} which is a factor $\sim 3.6$ less than
is predicted by the most detailed theoretical calculations,
$9.5_{-1.4}^{+1.2}$ SNU.\cite{BPBJCD,BP95}  
A SNU is a convenient unit to describe the
measured rates of solar neutrino experiments: $10^{-36}$ interactions
per target atom per second. 
Most of the predicted rate in the chlorine experiment is
from the rare, high-energy $^8$B neutrinos, although the $^7$Be
neutrinos are also expected to contribute significantly.  According to
standard model calculations, the $pep$ neutrinos and the CNO neutrinos 
(for simplicity not discussed here)
are expected to contribute less than 1 SNU to the
total event rate.

This discrepancy between the standard model calculations 
and the observations for the
chlorine experiment was, for more than two decades, the only solar
neutrino problem. I shall refer to the chlorine disagreement 
as the ``first'' solar neutrino
problem.

\subsection{Problem~2. Incompatibility of Chlorine and Water (Kamiokande) 
Experiments}

The second solar neutrino problem results from a comparison of the 
measured event rates in the chlorine experiment and in the Japanese
water Cerenkov experiment,  Kamiokande.  The water experiment detects
higher-energy neutrinos, those with energies  above $7$ MeV,
by neutrino-electron scattering: $\nu ~+~e ~\longrightarrow
  \nu ~+~e.$   According to the standard solar model, 
\hbox{$^{8}$B} beta decay is the only important 
source of these higher-energy neutrinos. 

The Kamiokande experiment shows that the observed neutrinos come from
the sun. 
The  electrons that are scattered by the incoming neutrinos recoil
predominantly  in the direction of the sun-earth vector;  the
relativistic electrons are observed by the 
Cerenkov radiation they produce in the water detector.

In addition, the Kamiokande 
experiment measures
the energies of individual scattered electrons and 
provides information about the energy spectrum of the incident 
solar neutrinos. The observed 
spectrum of electron recoil energies 
is consistent with that expected from $^8$B neutrinos.
However,  small angle scattering of the recoil  electrons in the water
prevents the angular distribution from being determined well on an
event-by-event basis, which limits  the constraints the experiment
places on the incoming neutrino energy  
spectrum.

The event rate in the Kamiokande experiment is
determined by the same high-energy $^8$B neutrinos that are expected,
on the basis of the combined standard model,
to dominate the event rate in the chlorine experiment.
 Solar physics changes   
the shape of the \hbox{$^{8}$B} neutrino spectrum by only 1 part
in $10^5$ (see Ref.~\citenum{Bahcall91}). 
Therefore, we can calculate the rate in the chlorine experiment that
is produced by  
the \hbox{$^{8}$B} neutrinos observed
in the Kamiokande experiment (above 7  MeV).
This partial (\hbox{$^{8}$B}) rate in the chlorine experiment 
is $3.2 \pm
0.45$ SNU, which exceeds the total observed chlorine 
rate of $2.55 \pm 0.25$ SNU. 

Comparing the rates of the
Kamiokande and the chlorine experiments, one finds that the
best-estimate net
contribution to the chlorine experiment from the $pep$, $^{7}$Be, and CNO
neutrino sources is negative: $-0.66 \pm 0.52$ SNU.
 The standard model calculated rate from $pep$, $^7$Be, and CNO neutrinos is
1.9~SNU.  The apparent incompatibility of the chlorine
and the Kamiokande
experiments is the ``second'' solar neutrino problem.
The inference that is most often made from this comparison is that the
energy spectrum of ${\rm ^8B}$ neutrinos is changed from the standard
shape by physics not included in the simplest version of the standard
electroweak model.

\subsection{Problem~3. Gallium Experiments: No Room for $^{7}$Be Neutrinos}
\label{galliumproblem}

The results of the 
gallium experiments, GALLEX and SAGE,
constitute the third solar neutrino problem.
The average observed  rate in these two experiments is $70.5 \pm 7$ 
SNU, which
is fully accounted for in the standard model by the 
theoretical  rate of $73$ SNU
that is calculated to come from the basic $pp$ and $pep$ neutrinos
(with only a 1\% uncertainty in the standard solar model $pp$ flux).
The \hbox{$^{8}$B} neutrinos, which are observed above $7.5$ MeV 
in the Kamiokande experiment, must also contribute to the gallium
event rate. 
Using the standard shape for the spectrum of ${\rm ^8B}$
neutrinos and normalizing to the rate observed in Kamiokande, 
${\rm ^8B}$ contributes  
another $7$ SNU,
unless something happens to the lower-energy  neutrinos
after they are created in the sun. (The predicted contribution is
16~SNU on the basis of the standard model.)  
Given the measured rates in the
gallium experiments, there is no room for the additional $34 \pm 4$
SNU that
is expected\cite{bahcall94} 
from $^{7}$Be neutrinos on the basis of 
standard solar models. 

The seeming exclusion of everything but $pp$ neutrinos in the gallium
experiments is the ``third'' solar neutrino problem.  This problem is
essentially independent of the  previously-discussed solar
neutrino problems, since this third problem  depends strongly 
upon the $pp$ neutrinos,  which  are not observed in
the other experiments. Moreover, the calculated 
$pp$  neutrino flux is approximately
 independent of solar models since it is closely related to the total
luminosity of the sun. 

The missing $^7$Be neutrinos cannot be
explained away by any change in solar physics. The \hbox{$^{8}$B}
neutrinos that are observed in the Kamiokande experiment are produced
in competition with the missing $^7$Be neutrinos; 
the competition is between electron capture on $^7$Be versus
proton capture on $^7$Be.
Solar model
explanations that reduce the predicted ${\rm ^7Be}$ flux 
generically reduce much more, too much,
 the predicted ${\rm ^8B}$ flux.

\begin{figure}[t]
\centerline{\psfig{figure=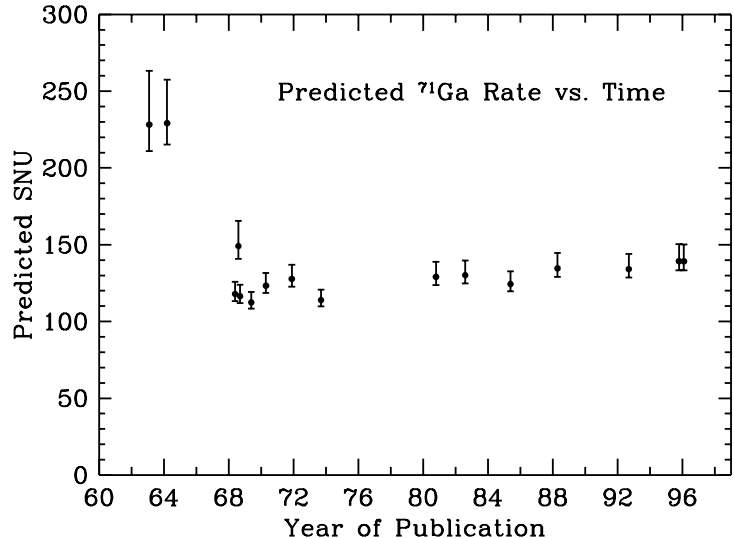,width=3.8in}}
\caption[]{\baselineskip=10pt\footnotesize Predicted Solar Neutrino Gallium Event Rate Versus 
Year of Publication.\cite{gallium}
The figure shows the event rates for all of the
standard 
solar model calculations that my colleagues and I have
published.  
The cross 
sections  from the recent paper by Bahcall\cite{gallium} 
have been used in all cases to convert the calculated neutrino fluxes to 
predicted capture rates.  
The estimated $1\sigma$ uncertainties that are shown reflect just the 
uncertainties in the cross sections that are calculated
 in Ref~\citenum{gallium}.
For the 35 years over which we have been calculating standard
solar model neutrino fluxes, the historically lowest value (fluxes 
published in 1969)
corresponds to $109.5$~SNU.  
This lowest-ever value is $5.6\sigma$ greater than the
combined GALLEX and SAGE experimental result.
If the points prior to 1992 are increased by 
\hbox{$11$~SNU} to correct for diffusion (this was not done in the
figure), 
then all of the standard model 
theoretical capture rates since 1968 through 1997 lie in the range 
$120$ SNU to $141$ SNU. 
\protect\label{gahistory}}
\end{figure}

The flux of $^7$Be neutrinos, $\phi({\rm ^7Be})$, is independent of
measurement uncertainties in  the
cross section for the nuclear reaction
${\rm ^7Be}(p,\gamma)^8$B; the  cross
section for this proton-capture  reaction is  the most uncertain
quantity that enters  in an important way in the solar
model calculations.  The flux of $^7$Be neutrinos depends upon the
proton-capture
reaction only through the ratio
\begin{equation}
\phi({\rm ^7Be}) ~\propto~ {{R(e)} \over {R(e) + R(p)}} ,
\label{Beratio}
\end{equation}
where $R(e)$ is the rate of electron capture by $^7$Be nuclei and
$R(p)$ is the rate of proton capture by $^7$Be.  With standard
parameters, solar models yield $R(p) \approx 10^{-3} R(e)$.
Therefore, one would have to increase the value of
the ${\rm ^7Be}(p,\gamma)^8$B cross section
by more than two orders of magnitude over the current best-estimate
(which has an estimated uncertainty of \hbox{$\sim$  10\%}) in order to affect
significantly the calculated $^7$Be solar neutrino flux.
The required change in the nuclear physics cross section
 would also  increase the predicted neutrino event
rate by more than a factor of 100 in the Kamiokande experiment, making that
prediction completely inconsistent with what is observed.
(From time to time, papers have been published claiming to solve the
solar neutrino problem by artificially changing the rate of the $^7$Be
electron capture reaction.
Equation~(\ref{Beratio}) shows that the flux of $^7$Be neutrinos
is independent of the rate of the electron capture reaction
to an accuracy of  better than 1\%.)

Figure~\ref{gahistory} shows the event rates for gallium
solar neutrino experiments that are predicted by all of the standard
solar model calculations that my colleagues and I have published in
the 35 years, 1962-1997, in which we have been calculating solar
neutrino fluxes.  The historically lowest values (fluxes published in
1969) corresponds to 109.5 SNU, 5.6 sigma greater than the combined
GALLEX and SAGE experimental result.  If the predictions prior to 1992
are increased by 11 SNU to correct for diffusion (this was not done in
the figure, but is required by helioseismological measurements, see
below), then all of the standard model theoretical capture rates since
1968 lie in the range 120 SNU to 141 SNU.  The solar model predictions
for the gallium experiment are robust!

\subsection{The bottom line}

If we adopt the combined standard model, 
Figure~\ref{compare} displays three solar neutrino problems: the 
smaller than predicted absolute event rates in the chlorine and
Kamiokande experiments, the incompatibility of the chlorine and
Kamiokande experiments, and the very low rate in the gallium
experiment (which implies the absence of $^7$Be neutrinos although $^8$B
neutrinos are observed). 

I conclude that either: 1) 
at least three of the four pioneering solar neutrino
experiments (the two gallium experiments plus either chlorine or
Kamiokande) 
 have yielded misleading results, or 2) physics beyond the minimal standard
electroweak model is required to change the neutrino energy spectrum (or
flavor content) after the neutrinos are produced in the center of the sun.

\section{Comparison with Helioseismological\\ Measurements}
\label{helioseismology} 

Helioseismology has recently sharpened
the disagreement between observations and the predictions of 
solar models 
with standard (non-oscillating) neutrinos.
The helioseismological measurements 
demonstrate that the sound speeds predicted by 
standard solar models agree with extraordinary precision with the
sound speeds of the sun inferred from helioseismological
measurements.\cite{Basu96a,Basu96b}  Because of the precision of this
agreement, I am convinced that standard solar models cannot be in
error by enough to make a major difference in the solar neutrino
problems.

I will report here on some work that Marc Pinsonneault,
Sarbani Basu, J{\o}rgen Christensen-Dalsgaard, and I have done recently 
which demonstrates the precise
agreement between the sound speeds in standard solar models and
the sound speeds inferred from helioseismological 
measurement.\cite{BPBJCD}

The square of the sound speed satisfies  $c^2 \simpropto T/\mu$, 
where $T$ is temperature and $\mu$ is mean molecular weight.
The sound speeds in the sun are determined from helioseismology to
a very high accuracy, better than 
 $0.2$\% rms throughout nearly all the sun.
Thus even tiny fractional  errors in  the model  
values of $T$ or $\mu$
would produce measurable discrepancies  in the precisely determined
helioseismological sound speed
\begin{equation}
{\delta c \over c}  \simeq 
{1 \over 2} \left({\delta T \over T}  - {\delta \mu \over \mu }\right)
\; .
\label{deltac}
\end{equation}
The numerical 
agreement 
between standard predictions and helioseismological 
observations, which I will discuss in the following remarks, 
 rules out   solar models with 
temperature or mean molecular weight 
profiles that differ significantly from standard profiles.
In particular, the helioseismological data essentially 
rule out solar models in which
deep mixing has occurred (cf.~PRL paper\cite{BPBJCD}) 
and argue against solar models in which the
subtle effect of particle diffusion--selective sinking
of heavier species in the sun's gravitational field--is not included.

\begin{figure}[t]
\centerline{\psfig{figure=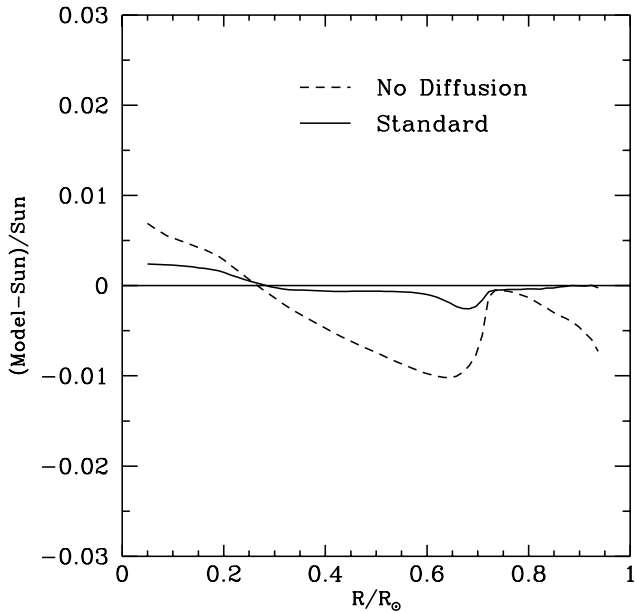,width=3.3in}}
\caption[]{\baselineskip=10pt\footnotesize Comparison of sound speeds predicted by different
standard solar models with the
sound speeds measured by  helioseismology.\cite{BPBJCD}  
There are no free parameters in the models. 
The figure shows the
fractional difference, $\delta c/c$, 
between the predicted model sound speed and the 
measured\cite{Basu96a,Basu96b}
solar values as a function of radial position in the sun
($R_\odot$ is the solar radius).
The dashed line refers to a model\cite{BP95} in which
diffusion is neglected  and 
the dark line represents a standard model which includes 
diffusion and recent improvements in the 
OPAL equation of state and opacities.\cite{BPBJCD}\label{fig:one}}
\end{figure}

Figure~\ref{fig:one} compares the  sound speeds computed
from two different solar models with the values
inferred\cite{Basu96a,Basu96b} from the helioseismological measurements.
The 1995, no diffusion, standard model of 
Bahcall and Pinsonneault (BP)\cite{BP95}
 is represented by
the dotted line; 
the dark line represents our  best solar model\cite{BPBJCD}
 which includes
recent improvements in the OPAL equation of state and
opacities, as well as helium and heavy element diffusion.
For the standard model with diffusion,
the rms discrepancy  
between  predicted and measured sound speeds
is  $0.1$\% (which is probably due in part to systematic uncertainties in the
data analysis that produced the solar sound speeds).

Figure~\ref{fig:one} shows that the discrepancies with the No
Diffusion model are as large as 
$1$\%.  
The mean squared discrepancy for the No Diffusion model is 22 times
larger than for the best model with diffusion, OPAL EOS.  If one
supposed optimistically that the No Diffusion model were correct, one
would have to explain why the diffusion model fits the data so much
better. 
On the basis of Figure~\ref{fig:one}, we conclude that otherwise
standard solar
models that do not include diffusion, such  as the model  of 
Turck-Chi\`eze 
and Lopez,\cite{TC93} 
are inconsistent with helioseismological observations.
This conclusion is consistent with earlier inferences based upon 
comparisons with  less complete
helioseismological data, including the fact that
the present-day surface helium abundance in a standard solar 
 model agrees with
observations only if diffusion is included.\cite{BP95}

Equation~\ref{deltac} and Figure~\ref{fig:one} imply that any changes
$\delta T/T$ from the standard model values of  temperature 
must be almost exactly canceled  by 
changes $\delta \mu/\mu$ in mean molecular weight.
In the standard solar model, $T$ and $\mu$ vary, respectively, by a factor
of $53$ and by $43$\% over the entire range for which $c$ has been
measured and by $1.9$ and $39$\% over the energy producing region.
It would be an extraordinary coincidence if nature chose $T$ and $\mu$
profiles that individually differ markedly from the standard model but
have the same ratio everywhere that they have  in the standard model.
There is no known 
reason why the large variation in $T$ should be finely tuned to the
smaller variation in $\mu$.  
In the absence of a cosmic conspiracy, I conclude 
that the fractional differences between the solar temperature and the model
temperature, $\delta T/T$, or the fractional differences between 
mean molecular weights, $\delta \mu/\mu$,
are of similar magnitude to $\delta c^2/c^2$, i.e. (using the larger 
rms error, $0.002$, for the solar interior),

\begin{equation}
\vert \delta T/T \vert, ~\vert \delta \mu/\mu \vert ~ \ltorder ~ 0.004 .
\label{inequality}
\end{equation}

How significant for solar neutrino studies 
is the agreement between observation and prediction
that is shown in Figure~\ref{fig:one}? 
The calculated neutrino fluxes 
depend upon the central
temperature of the solar model 
approximately as a power of the temperature, 
${\rm Flux} \propto T^n$, where  for standard models 
the exponent $n$ varies 
from $n \sim -1.1$ for
the  $pp$ neutrinos to  $n \sim +24$ for the $^8$B 
neutrinos.\cite{BU96}  
Similar temperature scalings are found for non-standard solar 
models.\cite{Castellani94,Castellani97} Thus,
 maximum temperature differences of
$\sim 0.2\%$ would produce changes in the different neutrino
fluxes of several percent or less, more than an order of magnitude less than 
required\cite{newphysics} to
ameliorate the solar neutrino problems discussed in Section~\ref{threeproblems}.

Helioseismology rules out all solar models with large amounts of
interior mixing (which homogenizes the mean molecular weight), 
unless finely-tuned compensating changes in the
temperature are made.  The mean molecular weight
in the standard solar model with diffusion varies monotonically 
from $0.86$ in the
deep interior to $0.62$ at the outer region of nuclear fusion 
($R = 
0.25 R\odot$) to $0.60$ near the solar surface.
Any mixing
model will cause $\mu$ to be constant and equal to
the average value in the mixed region.
At the very least, 
the region in which nuclear fusion occurs must  be mixed 
in order to 
affect significantly the calculated neutrino 
fluxes.\cite{Bahcall89,EC68,BBU68,SS68,Schatzman69}
Unless almost precisely canceling temperature changes are assumed,
solar models in which the nuclear burning region is mixed ($R \ltorder
0.25 R_{\odot}$)
will give maximum  differences, $\delta c$, between
the mixed and the 
standard model predictions, and hence between the mixed model
predictions and the observations, of order
\begin{equation}
{\delta c \over c} ~=~
{1 \over 2} \left({ {\mu - < \mu >} \over {\mu} }\right) ~\sim~ 7\%
~{\rm to}~ 10\%,
\label{maximum}
\end{equation}
which is inconsistent with Figure~\ref{fig:one}.

\section{``The Last Hope'': No Solar Model}
\label{lasthope}

The clearest way to see that the results of the four solar neutrino
experiments are inconsistent with the predictions of the minimal standard
electroweak model is not to use standard solar models at all in the
comparison with observations. 
This is what Berezinsky, Fiorentini, and
Lissia\cite{Berezinsky96} have termed ``The Last Hope'' for a solution
of the solar neutrino problems without introducing new physics.

Let me now explain how model independent tests are made.

Let $\phi_i(E)$ be the normalized shape of the neutrino energy
spectrum from one of the neutrino sources in the sun (e.g., 
$^8$B or $pp$ neutrinos). I have shown\cite{Bahcall91} that the shape
of the neutrino energy spectra that result from radioactive decays, 
 $^8$B, $^{13}$N, $^{15}$O, and $^{17}$F, are the same 
to $1$ part in $10^5$ as the laboratory shapes.  The $pp$ neutrino
energy spectrum, which is produced by fusion  has a slight dependence
on the solar temperature, which affects the shape by about $1$\%. 
The energies of the neutrino lines from $^7$Be and $pep$ 
electron capture reactions are
also only shifted slightly, by about  $1$\% or less, 
because of the thermal energies of particles in the solar core.  

Thus a test of the hypothesis that an arbitrary linear combination
of the normalized standard neutrino spectra,

\begin{equation}
\Phi(E) ~=~\sum_{i} \alpha_i \phi_i(E),
\label{arbitrary}
\end{equation}
can fit the results of the neutrino experiments is equivalent to a
test of minimal standard electroweak theory.  One can choose the
values of $\alpha_i$ so as to minimize the discrepancies with existing
solar neutrino measurements and ignore all solar model information
about the $\alpha_i$.
One can add a constraint to Equation~(\ref{arbitrary}) that embodies the
fact that the sun shines by nuclear fusion reactions that also produce
the neutrinos.  
The explicit form of this luminosity constraint is 
\begin{equation}
\frac{L_\odot}{4\pi r^2} = \sum_j \beta_j \phi_j~,
\label{explicit} 
\end{equation}
where the eight coefficients, $\beta_j$, are determined by laboratory
nuclear physics measurements and are  given in
Table~VI of the paper by Bahcall and Krastev.\cite{BaKr}

The first demonstration that the four pioneering experiments are
by themselves inconsistent with the assumption that nothing happens to
solar neutrinos after they are created in the core of the sun was by 
Hata, Bludman,
and Langacker.\cite{Hata94}  
They showed that the solar neutrino data available by late 1993 were
incompatible with any solution of Equations~(\ref{arbitrary}) and 
(\ref{explicit}) at the 97\% C.L. 

In the most recent and complete published 
analysis in which the neutrino fluxes
are treated as free parameters,
Heeger and Robertson\cite{HR96} showed that the data
presented at the Neutrino '96 Conference in Helsinki are  inconsistent
with Equations~(\ref{arbitrary}) and 
(\ref{explicit}) at the 99.5\% C.L.   Even if they omitted the
luminosity constraint, Equation~(\ref{explicit}), they found
inconsistency at the 94\% C.L. Similar results have been obtained by
Hata and Langacker.\cite{HL97} 

It seems to me that these demonstrations are so powerful and general
that there is very little point in discussing potential ``solutions''
to the solar neutrino problem based upon hypothesized non-standard
scenarios for solar models.

\section{Summary of the First Lecture}
\label{closing}

The combined predictions of the standard solar model and the 
minimal standard
electroweak theory disagree with the results of the four pioneering
solar neutrino experiments.  The disagreement persists even if the 
neutrino fluxes are treated as free parameters, without reference to 
any solar model.

The  solar model calculations are in excellent agreement with 
helioseismological measurements of the sound speed, providing
further support for the inference that something happens to the solar
neutrinos after they are created in the center of the sun.

Looking back on what was envisioned in 1964, I am astonished and
pleased with what has been accomplished. In 1964, it was not clear 
that solar neutrinos could be detected.  Now, they have been observed
in five different experiments (including the results reported for
Super-Kamiokande at this
School) and the theory of stellar energy generation by
nuclear fusion has been directly established. 
Moreover, helioseismology has confirmed to high precision predictions
of the standard solar model, a possibility that  also was not imagined
in 1964.
Particle
theorists have shown that solar neutrinos can be used to study
neutrino properties, another possibility that we did not envision in
1964.  Much of the interest in the subject now stems from the
unanticipated fact that 
the four pioneering experiments suggest that new neutrino physics may 
be revealed by solar neutrino measurements.
We shall discuss in the next lecture some of 
the possibilities for detecting unique signatures of
new physics with the powerful second generation of solar neutrino 
 experiments that are now beginning
to operate.  

\section{New Solar Neutrino Experiments}
\label{newexperiments}

I would like to begin this second lecture by listing the new solar
neutrino experiments.   
Table~\ref{nuexperiments} shows the new experiments 
that are operating, under construction, or are being
developed.  You have already heard a lot about these experiments in the
lectures by K. Martens.

\begin{table}[t]
\centering
\caption[]{New Solar Neutrino
Experiments\protect\label{nuexperiments}}
\begin{tabular}{l@{\extracolsep{-2pt}}cccc}
\noalign{\medskip}
\hline\hline
\noalign{\smallskip}
\multicolumn{1}{c}{Collaboration}&$\nu$'s&Detector&Technique&Beginning\\
&&&&Date\\
\noalign{\smallskip}
\hline
\noalign{\medskip}
Super-Kamiokande&${\rm ^8B}$&22.5 kt ${\rm H_2O}$&$\nu$-$e$
scattering&April '96\\
\noalign{\smallskip}
SNO&${\rm ^8B}$&1 kt ${\rm D_2O}$&abs., nc disint.&Early '98\\
\noalign{\smallskip}
GNO&$pp$, ${\rm ^7Be}$ +...&30--100 t Ga&radiochemical&Early '98\\
\noalign{\smallskip}
BOREXINO&${\rm ^7Be}$&100 t liquid&$\nu$-$e$ scattering&'99\\
&&scintillator\\
\noalign{\smallskip}
ICARUS&${\rm ^8B}$&600 t liquid Ar&$\nu_e$ abs., TPC&'99\\
\noalign{\smallskip}  
Iodine&${\rm ^7Be, ^8B}$,...&100 t iodine&radiochemical&'99\\
\noalign{\smallskip}
HELLAZ&$pp$, ${\rm ^7Be}$&gaseous He&$\nu_e$-scattering&Develop.\\
&&&(TPC)\\
\noalign{\smallskip}
HERON&$pp$, ${\rm ^7Be}$&liquid He&$\nu_e$-scattering&Develop.\\
&&&(superfluid, rotons)\\
\noalign{\medskip}
\hline
\end{tabular}
\end{table}

I only want to add a few summary words.  The 
Super-Kamiokande,\cite{Takita93,Totsuka96}
SNO,\cite{McD94} and
BOREXINO\cite{Arp92}
 experiments all detect the recoil electrons produced by the
neutrino interactions using Cerenkov detectors.  The radiochemical
experiments, GNO and Iodine (${\rm ^{127}I}$),\cite{haxton87,cleveland} 
detect neutrinos above a fixed
threshold ($0.23$ MeV for GNO and $0.67$ MeV for Iodine) by 
counting the chemically extracted 
radioactive product ($^{71}$Ge or $^{127}$Xe) in a small
proportional counter.
All of the
other experiments measure electronically 
energies associated with individual neutrino
events. Among experiments that will operate before the year 2000, only
GNO is sensitive to the low energy neutrinos from the fundamental $pp$
reaction and only BOREXINO can measure separately the flux of
neutrinos from $^7$Be electron capture, the crucial $^7$Be neutrino
line.  SNO is the only experiment listed that can measure the total
flux of neutrinos (of any flavor), which will be accomplished 
using the neutral current
disintegration of deuterium. The neutrino interaction cross sections are well known (typical
accuracy of order a few percent or better) for all of the detectors
except $^{127}$I.

\section{Why do physicists care about solar neutrinos?}
\label{whyphysicists}

Solar neutrinos are of interest to physicists 
because they can be used to perform  unique particle physics 
experiments. Many physicists believe that  solar 
neutrino experiments may in fact 
have already provided strong hints that at least one neutrino
type has a non-zero mass and that electron flavor (or the number of
electron-type neutrinos) may not be conserved.

For some of the theoretically most interesting
ranges of masses and mixing angles, 
solar neutrino experiments are more  sensitive 
tests for neutrino  transformations in flight  than
experiments that can be carried out with laboratory sources.  The 
reasons for this exquisite sensitivity are:
1) the great distance between the beam source (the solar
interior) and the detector (on earth); 2) the relatively low energy 
(MeV) of
solar neutrinos; and 3) the enormous path length
of matter ($\sim 10^{11} {\rm gm~ cm^{-2}}$) that neutrinos must pass
through on their way out of the sun.  

One can quantify the sensitivity of solar neutrinos  relative to
laboratory experiments by considering the proper time that would
elapse for  a
finite-mass neutrino   in  flight between the point of
production and the point of detection.  The elapsed proper time 
is a measure of
the opportunity that a neutrino has to transform its state and is
proportional to the ratio, $R$, of path length divided by energy:
\begin{equation}
{\rm Proper~Time} ~\propto~R =~ {\rm {Path~Length} \over {Energy} }.
\label{proper}
\end{equation}

Future accelerator experiments 
with multi-GeV neutrinos may reach a
sensitivity of $R = {\rm 10^2~km~GeV^{-1}}$.
Reactor  experiments  have already reached a level of
sensitivity of  
$R = {\rm 10^{2.5}~km~GeV^{-1}}$ for neutrinos with 
MeV energies\cite{chooz} and are
expected to improve to ${\rm 10^4~km~GeV^{-1}}$.
Solar neutrino experiments, because of the enormous distance between
the source (the interior of the sun) and the detector (on earth) and the
relatively low energies ($1$ MeV to $10$ MeV) of solar neutrinos
involve much larger values of 
 neutrino proper time, 
\begin{equation}
R({\rm solar}) = { {{10^8} \over {10^{-3}} } ~{\left(\rm {km}
\over {GeV} \right) } ~\sim~10^{11}
~{\left(\rm {km}
\over {GeV} \right) } }.
\label{Rsolar}
\end{equation}

Because of the long proper time that is available to a neutrino to
transform its state,  solar neutrino experiments
are sensitive to very small neutrino masses that can cause  neutrino
oscillations in vacuum.  Quantitatively, 
\begin{equation}
m_\nu( {\rm solar~level~of~ sensitivity}) \sim 
10^{-6} {\rm eV~to} ~10^{-5} {\rm eV}~~
{\rm \left(vacuum~oscillations\right) },
\label{vacuumrange}
\end{equation}
provided the electron neutrino that is created by beta-decay contains
appreciable portions of at least two different neutrino mass eigenstates
(i.e., the neutrino mixing angle is relatively large).
Direct laboratory experiments have achieved a sensitivity  to electron
neutrino masses of order a few eV.  Over the next several years, the
sensitivity of the laboratory experiments may be improved by an
order of magnitude or more.

Resonant neutrino oscillations, which may be 
induced by neutrino interactions with electrons
in the sun (the famous Mikheyev-Smirnov-Wolfenstein, MSW,\cite{msw} effect), can occur even if the electron
neutrino is almost entirely composed of one neutrino mass eigenstate
(i.e., even if 
the mixing angles between $\nu_e$ and $\nu_\mu$ and between $\nu_e$ and 
$\nu_\tau$ neutrinos are tiny). 
Standard solar models indicate that 
the sun has a high central density, $\rho({\rm central}) \sim 1.5
\times 10^2~{\rm gm~cm^{-3}}$, which 
allows even very low energy ($ < 1$ MeV) electron neutrinos
to be resonantly converted to the more difficult to detect $\nu_\mu$ or 
$\nu_\tau$ neutrinos
by the MSW effect. Also, the column density
of matter that neutrinos must pass through is large: $\int \rho dr
\approx 2 \times 10^{11}~{\rm gm~cm^{-2}}$.
The corresponding 
parameters for terrestrial, long-baseline experiments are:\  a 
typical density of
${\rm 3~ gm~cm^{-3}}$, and an  obtainable column density of  
$\sim 2 \times 10^{8}~{\rm gm~cm^{-2}}$.

Given the above solar parameters,
the planned and operating solar
neutrino experiments are sensitive to neutrino masses in the range

\begin{equation}
10^{-4}~{\rm eV} ~\ltorder~ m_\nu ~\ltorder~ 10^{-2}~{\rm eV}, 
\label{MSWrange}
\end{equation}
via matter-induced resonant oscillations (MSW effect).

The range of neutrino masses given by Equation~(\ref{vacuumrange}) and
Equation~(\ref{MSWrange}) is included in the range of neutrino masses that
are suggested by attractive particle-physics generalizations of the minimal
standard electroweak model, including left-right symmetry,
grand-unification, and supersymmetry.

Both vacuum neutrino oscillations and matter-enhanced neutrino
oscillations can change electron neutrinos to the more difficult
to detect muon or tau neutrinos (or even, in principle, to sterile
neutrinos).  In addition, the likelihood that a
neutrino will have its flavor  changed
may depend upon its energy, thereby affecting the shape of
the energy spectrum of the surviving electron neutrinos.  Future
solar neutrino experiments will
measure the shape of the recoil 
electron  energy spectrum (produced via charged current absorption and by
neutrino-electron scattering)
and will also measure
the ratio of the number of
electron neutrinos to the total number of solar neutrinos (via
neutral current reactions).
These measurements, of the spectrum
shape and of the ratio of electron-type
 to total number of neutrinos, will  test
the simplest version of the minimal standard electroweak model in which
neutrinos are massless and do not oscillate. These tests
are independent of
solar model physics.

For simplicity in the following discussions of both MSW and vacuum
oscillations, I will assume that only two types of neutrinos are
mixed.  A richer set of solutions can be obtained if this assumption
is dropped (see, e.g., the lectures by H. Harari at this summer
school or the paper by Folgi {\it et al.},\cite{fogli97a} both of which contain a useful
set of further references).

\section{Allowed MSW solutions}
\label{allowedmsw}

The most popular neutrino physics solution,   the 
Mikheyev-Smirnov-Wolfenstein (MSW) effect,\cite{msw} 
predicts several characteristic phenomena that are not expected if 
 minimal standard electroweak theory is correct.
The MSW effect explains solar neutrino observations as the result of
conversions in the solar interior of  $\nu_e$ 
produced in nuclear reactions
to the more difficult to detect $\nu_\mu$ or $\nu_\tau$.

Potentially decisive signatures of new physics that 
are suggested by the MSW effect 
include observing that the sun is brighter in
neutrinos at night (the `earth regeneration effect'), 
\cite{earthreg,BaltzW,BaltzW2}
detecting distortions in the 
incident solar neutrino energy spectrum,\cite{elspectJ}
and observing that the flux of all types of neutrinos exceeds the flux
of just electron  neutrinos.\cite{chen}
A demonstration that any  of these 
 phenomena exists would provide evidence for physics beyond the
minimal standard electroweak model.
I shall discuss in the next section 
the possibilities for detecting each of these
signatures within the context of ``The Search for Smoking Guns.''

Including the earth regeneration effect, Plamen Krastev and I\cite{BK97} 
 have calculated
the expected one-year average event rates as functions of the neutrino
oscillation parameters, $\Delta m^2$ (the difference in squared
neutrino masses), and $\sin^22\theta$ (where $\theta$ is 
the mixing angle between $\nu_e$ and the mass eigenstate that $\nu_e$
most resembles),
for all four
operating experiments which have published results from their
measurements of solar neutrino event rates. Specifically, the
experiments included are the Homestake chlorine
experiment, Kamiokande, GALLEX and SAGE. 
 We take into account the known threshold and 
cross-sections for each detector. In the case of 
Kamiokande, we also take into account
the known energy resolution (20\%, $1\sigma$, at electron energy 10 MeV) and
trigger efficiency function.\cite{Kam} For similar calculations and
related references, see, e.g., the papers by  Maris and
 Petcov\cite{maris} and Lisi and Montanino.\cite{lisi97}

We first calculate the one year average survival probability,
$\bar{P}_{SE}$, for a large number of values of $\Delta m^2$ and
$\sin^22\theta$.  Then we
compute the corresponding one year average event rates in each
detector. We perform a $\chi^2$ analysis taking into account
theoretical uncertainties and experimental errors.
 We obtain allowed regions in $\Delta m^2$ -
$\sin^22\theta$ parameter space by finding the minimum $\chi^2$ and
plotting contours of constant $\chi^2 = \chi^2_{min} + \Delta\chi^2$
where $\Delta\chi^2 = 5.99$ for 95\% C.L. and 9.21 for 99\% .

The best fit is obtained for the small mixing angle (SMA) solution:

\begin{eqnarray}
\label{SMAall}
\Delta m^2  &=&  5.0\times 10^{-6} {\rm eV}^2 ,\nonumber\\ 
\sin^22\theta  &=&  8.7\times 10^{-3} ,
\end{eqnarray}
which has a $\chi^2_{\rm min} = 0.25$.  
There are two more local minima of $\chi^2$.  The best fit for the
well known large mixing angle (LMA) solution occurs at

\begin{eqnarray}
\label{LMAall}
\Delta m^2  &=&  1.3\times 10^{-5} {\rm eV}^2 ,\nonumber\\
\sin^22\theta  &=& 0.63 ,
\end{eqnarray}
with $\chi^2_{\rm min} = 1.1$.  There is also a less probable
solution,\cite{krastev93,BaltzW3} which we refer to as 
the LOW solution (low probability, low
mass), at

\begin{eqnarray}
\label{LOWall}
\Delta m^2 & = &1.1\times 10^{-7} {\rm eV}^2 ,\nonumber\\
\sin^22\theta & = & 0.83 .
\end{eqnarray}
with $\chi^2_{\rm min} = 6.9$. 
 The LOW solution is acceptable only
at 96.5\% C.L.

\begin{figure}
\centerline{\psfig{figure=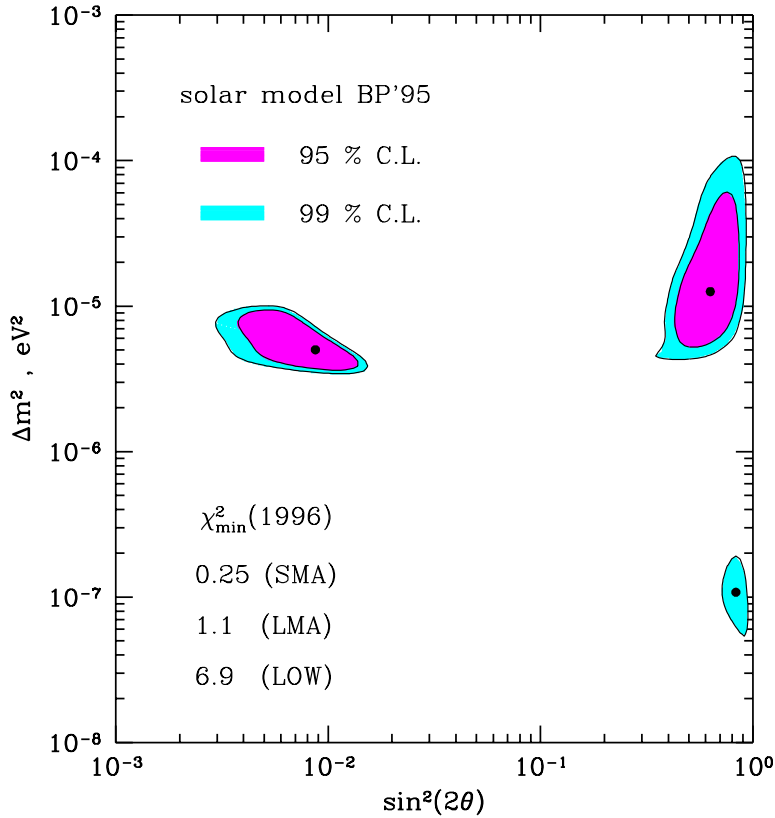,width=4in}}
\vglue-.3in
\caption[]{\baselineskip=10pt\footnotesize 
Allowed MSW solutions with regeneration.\cite{BK97}
The allowed regions are shown for the
neutrino oscillation parameters $\Delta m^2$ and $\sin^22\theta$.
The C.L. for the outer regions is 99\% 
 and the  C. L for the inner regions 
is 99\% (only applies to the LMA and SMA
solutions).
The data used here are
from the Homestake 
chlorine,\cite{cleveland98,homestake} Kamiokande,\cite{Suzuki95,kamioka}
GALLEX,\cite{Ansel95,GALLEX} 
and SAGE\cite{Abdur94,SAGE} 
experiments.  The solar model used is the best standard model of
Bahcall and Pinsonneault (1995) with helium and heavy element
diffusion.\cite{BP95} 
The points where $\chi^2$ has a  local minimum are indicated
by a circle. 
\protect\label{allowed}}
\end{figure}
 
Figure~\ref{allowed} shows the allowed regions in  
the plane defined by $\Delta m^2$ and $\sin^2 2 \theta$.  The C.L. 
is 95\% for
the allowed regions of the SMA and LMA solutions and 99\% for the LOW
solution. 
The black dots within
each allowed region indicate the position of the local best-fit point
in parameter space.
The results shown in Fig.~\ref{allowed}
were calculated using the predictions of
the 1995 standard solar model of Bahcall and
Pinsonneault,\cite{BP95} 
which includes helium and heavy element diffusion;
the shape of the allowed contours depends only slightly upon the 
assumed solar model (see Fig.~1 of (Ref.~\citenum{BaKr}).

The predicted $\nu-e$ scattering rates for the ${\rm 0.86~MeV~^7Be}$
line (which will be studied by BOREXINO\cite{Arp92}) 
relative to the Bahcall and Pinsonneault 1995 standard model \cite{BP95}
are: $0.22^{+0.18}_{-0.00}$ (SMA), $0.54^{+0.17}_{-0.16}$ (LMA), and
$0.54^{+0.08}_{-0.07}$ (LOW).  The SMA and LMA ranges correspond to
95\% C.L. and the LOW range is 99\% C.L.

\begin{figure}[t]
\centerline{\psfig{figure=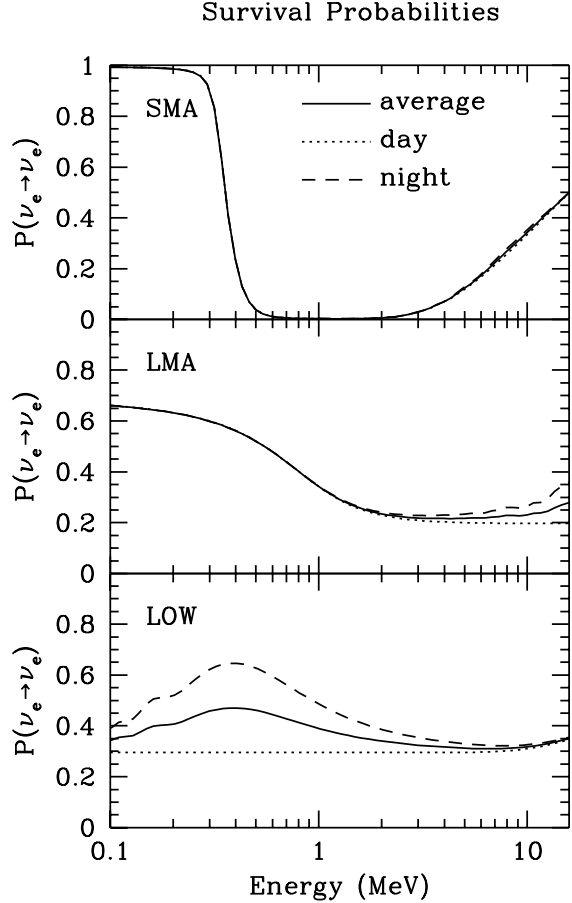,width=3in}}
\caption[]{\baselineskip=10pt\footnotesize Survival probabilities for MSW 
solutions.\cite{BK97}
The figure presents the survival probabilities for a $\nu_e$ created
in the sun to remain a $\nu_e$ upon arrival at the 
 earth.  The best-fit MSW solutions including regeneration in the
earth are described in
in the text.
The full line refers to 
the average
survival probabilities computed  taking
into account regeneration in the earth and the dotted line refers to
calculations for the day-time 
that do not include regeneration. 
The dashed line includes regeneration at night.
There are only slight
differences between the computed regeneration probabilities for
the detectors located at the positions of Super-Kamiokande, 
SNO and the Gran Sasso
Underground Laboratory.
\protect\label{survival}}
\end{figure}

Figure~\ref{survival} compares the  computed survival probabilities
for the day (no regeneration), the night (with regeneration), and the 
annual average.
These results show that there are day-night shifts in the neutrino
energy spectrum as well as in the total rate, i.e., the shape of the
effective 
$\nu_e$ energy spectrum depends upon the solar zenith angle. 
The  results in the figure refer to a detector at the location of
Super-Kamiokande, but the differences are very small between the survival
probabilities at the positions of Super-Kamiokande, SNO, and the Gran
Sasso Underground Laboratory. 

\section{Vacuum Neutrino Oscillations}
\label{vacuum}

Historically, neutrino oscillations in vacuum\cite{pontecorvo} 
was the first suggested particle-physics 
solution to what was then the single ``solar neutrino problem'', the
fact that the rate  of occurrence of neutrino events in the chlorine detector 
was smaller than predicted by standard solar models and the assumption
that nothing happened to the neutrinos after they were produced.

\begin{figure}[tb]
\centerline{\psfig{figure=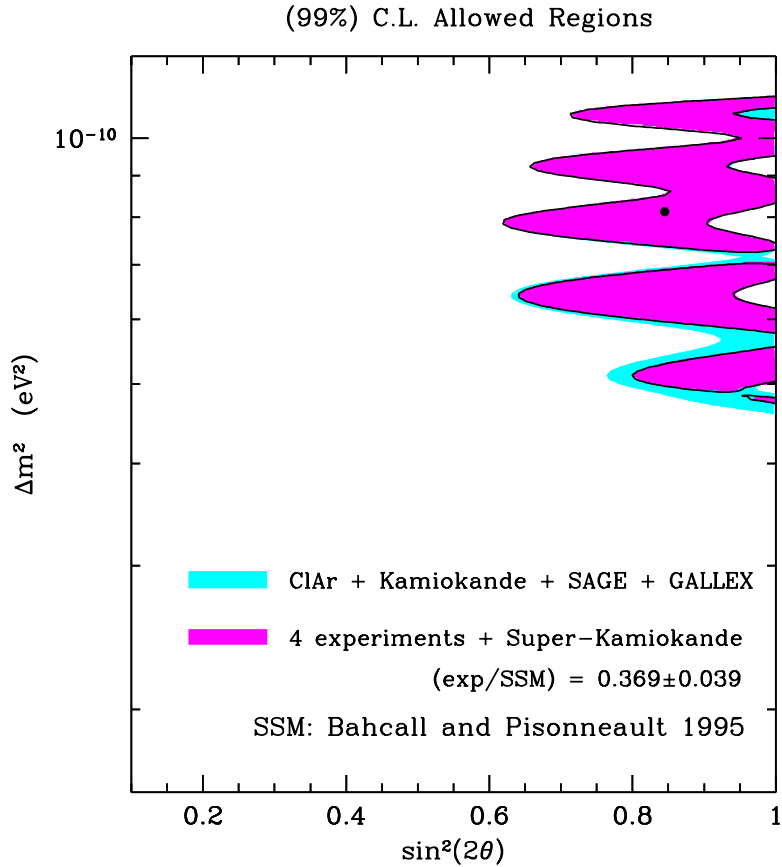,width=4in}}
\caption[]{\baselineskip=10pt\footnotesize Allowed vacuum oscillation solutions.
The allowed regions are shown  at the 99\% C.L. for the
neutrino oscillation parameters $\Delta m^2$ and $\sin^22\theta$.
The data used here are
from the Homestake 
chlorine,\cite{cleveland98,homestake} Kamiokande,\cite{kamioka}
GALLEX,\cite{GALLEX} 
and SAGE\cite{SAGE} 
experiments, and a preliminary report from the Super-Kamiokande
experiment (see lectures in this volume).   The solar model used is the best standard model of
Bahcall and Pinsonneault (1995) with helium and heavy element
diffusion.\cite{BP95} 
The point where $\chi^2$ has a  local minimum are indicated
 by a circle.  This figure was prepared by P. Krastev.
\protect\label{vacuumoscfig}} 
\end{figure}

Figure~\ref{vacuumoscfig} shows the allowed range of solutions for
vacuum oscillations, taking account of the four pioneering solar
neutrino experiments and preliminary results from Super-Kamiokande. 
This figure was prepared by Plamen Krastev as part of our ongoing
collaboration with Alexei Smirnov.  The calculations were performed
using the same data and methods described in the previous section
 in connection
with the discussion of allowed MSW solutions.

\section{The Search for Smoking Guns}
\label{smoking}
The new generation of solar neutrino experiments will carry out tests
of minimal standard electroweak theory that are independent of solar models.
These experiments are designed to have the capabilities of detecting
unique signatures of new physics, such as finite neutrino mass and
mixing of neutrino types.  For brevity, I shall refer to tell-tale
evidences of new physics as ``smoking guns.''

I will base the discussion of MSW smoking guns on three papers by 
Plamen Krastev, Eligio Lisi, and myself.\cite{BK97,snoJE,BKL97}
Similar papers have been written by other authors (see, for example, 
references in
our papers), but I use our work
here because I am most familiar with the details of what we did and
because I have easy access to our figures. Our results are generally
more pessimistic (indicate less sensitivity to new physics) than most
of the other published works. This is because we have included
estimates of the systematic uncertainties in our simulations, whereas
most other workers have only included statistical errors. I will base
the discussion of vacuum oscillations on the papers by Fogli, Lisi,
and Montanino,\cite{Fogli97} and Krastev and Petcov.\cite{KP93}

I will begin by describing in outline form  how we have determined preliminary
estimates of the likely sensitivities of the new solar neutrino
experiments.  Given the data from the four pioneering experiments
(Homestake chlorine, Kamiokande, GALLEX, and SAGE), we determine the
best-fit parameters, and the range of allowed solutions (at a
specified C.L.), for a given model of new neutrino physics (e.g.,
vacuum neutrino oscillations or the MSW effect). Then we 
calculate the expected rates in 
the new experiments (Super-Kamiokande,\cite{Takita93,Totsuka96}
 SNO,\cite{McD94} BOREXINO,\cite{Arp92} ICARUS,\cite{revol} 
HERON,\cite{lanou} or HELLAZ\cite{seguinot}) 
for all values of the new neutrino physics
parameters that are suggested by the pioneering experiments.
We take account of the characteristics of the new detectors that
the experimental collaborations say are expected. For example, 
we include, in addition to statistical errors,
the errors in the absolute energy determination of recoil electrons,
the width and uncertainty of the energy resolution function,
and the efficiency of detection, as well as uncertainties in the input
theoretical quantities (like the shape of the intrinsic 
neutrino energy spectrum and uncertainties in neutrino interaction
cross sections).
We do not include the effects of background events, because
the size of the backgrounds are not yet well known.  

Full Monte Carlo simulations of the detectors will be necessary to
determine accurately the sensitivities of each of the new
experiments. These detailed simulations can only be done by the
relevant experimental collaboration, since only the
collaboration will
have all the data required to make a realistic representation of how
the detector operates.  

In a survey of sensitivities, it is convenient to use 
the first two moments of the  observable distributions predicted by
different neutrino scenarios (e.g., the first two moments
 of the recoil electron energy spectrum or the zenith angle of the sun at the 
time of occurrence of neutrino
events). My colleagues and I have shown by detailed analyses that the
first two moments of the recoil energy spectrum or the solar zenith
angle contain most of the important information.

\subsection{Does the Sun Appear Brighter at Night\\ in Neutrinos?}
\label{daynight}

The MSW solution of the solar neutrino problems requires that  
electron neutrinos produced in nuclear reactions in the center of
the sun are converted to muon or tau neutrinos by interactions with
solar electrons on their way from the interior of the sun 
to the detector on earth.  
The conversion in the sun is primarily a resonance phenomenon, 
which---for each neutrino energy---occurs at a specific 
density (for a specified neutrino mass difference).

During day-time, the higher-energy neutrinos
arriving at earth are mostly $\nu_\mu$ (or $\nu_\tau$) with some
admixture of $\nu_e$. At night-time,  neutrinos must pass
through the earth in order to reach the detector.
As a result of traversing the earth,    the 
fraction of the more easily 
detected $\nu_e$ increases
because of the conversion of $\nu_\mu$ (or $\nu_\tau$) to 
$\nu_e$  by neutrino oscillations.  For the small mixing angle MSW
solution, 
interactions with electrons in the earth increase the effective 
mixing angle and enhance the conversion process.  For the large mixing
angle MSW solution, the conversion of $\nu_\mu$ (or $\nu_\tau$) to 
$\nu_e$ occurs by oscillations that are only slightly enhanced over
vacuum mixing.
This process of increasing in the earth 
the fraction of the neutrinos that are
$\nu_e$ is   called the ``regeneration effect'' and has  the
opposite effect to 
 the conversion of $\nu_e$ to $\nu_\mu$ (or $\nu_\tau$) in
the sun. 

Because of the change of neutrino flavor  in the earth, the
MSW mechanism predicts that solar neutrino detectors 
should generally measure
higher event rates at night than during day-time.

The regeneration effect is an especially powerful diagnostic of new
physics since no
difference is predicted between the counting rates 
observed during
the day and at night 
(or, more generally, any dependence of the counting rate on the solar
zenith angle) by such popular alternatives
to the MSW effect  as vacuum oscillations,\cite{pontecorvo} 
magnetic moment
transitions,\cite{okun} or violations of the equivalence
principle.\cite{gasperini}

\begin{figure}[t]
\begin{minipage}[t]{5.75in}
\centerline{\psfig{figure=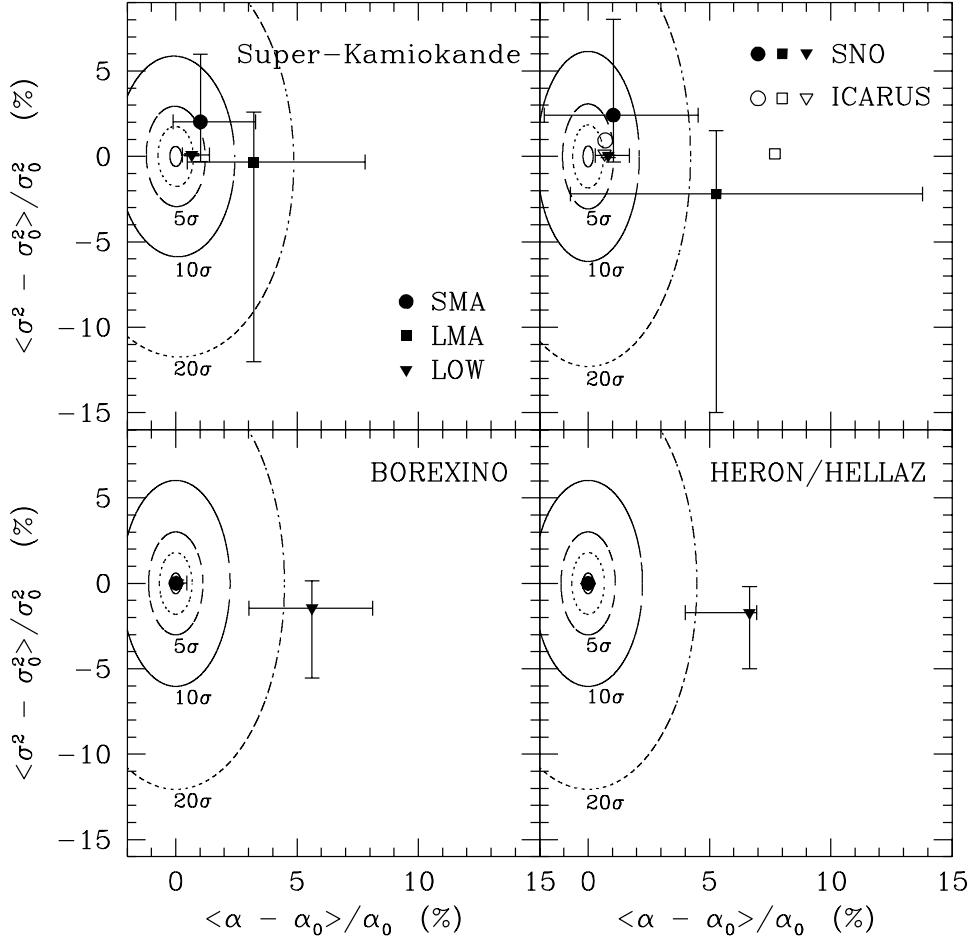,width=5in}}
\vglue.3in
\caption[]{\baselineskip=10pt\footnotesize How many sigmas?  The figure shows the sensitivity of 
Super-Kamiokande, SNO, ICARUS, BOREXINO and
HERON/HELLAZ to the regeneration effect.  Iso-sigma contours,
statistical errors only, 
delineate the
fractional percentage 
shifts of the first two moments of the angular distribution
of events for an assumed 30000 observed events.
Here $<\alpha>$ is the average solar zenith angle at the time of
occurrence of solar neutrino events and $\sigma$ is the dispersion
in the solar zenith angles.
For all but the ICARUS experiment,
the best-fit MSW solutions are indicated by black circles
(SMA), squares (LMA), and triangles (LOW);
the best-fit solutions are presented in Section~\ref{allowedmsw}.   The error
bars on the predicted moments correspond to $\Delta m^2$ and
$\sin^22\theta$ within allowed solution space  at 95\% C.L. (for
Super-Kamiokande, SNO, and ICARUS) or 99\% C.L. 
(BOREXINO and HERON/HELLAZ).
For ICARUS, we have indicated the best-fit solutions by a transparent
circle, square, or triangle. The best-fit SMA and LOW solutions for
ICARUS and the LOW solution for SNO are all three close together at
about $3\sigma$ from the no oscillation solution. In order to avoid
too much crowding in the figure, we have not shown the theoretical 
uncertainties for ICARUS. This figure is
 Figure 9 of Bahcall and Krastev.\cite{BK97} 
\protect\label{isosigma}}
\vglue2in
\end{minipage}
\end{figure}

Figure~\ref{isosigma}
summarizes the potential of the second generation of solar neutrino
experiments for discovering new physics via the earth regeneration
effect. The figure
displays iso-sigma ellipses, statistical errors only, in the plane of the
fractional percentage shifts of the first two moments, 
$\Delta \alpha/\alpha_0$ and $\Delta
\sigma^2/\sigma_0^2$. Here $<\alpha>$ is the average 
solar zenith angle at the time of
occurrence of solar neutrino events and $\sigma$ is the dispersion
in the solar zenith angles.

Assuming a total number of events of 30000, (which corresponds to 
$\sim 5$ years
of standard operation for Super-Kamiokande 
and $\sim 10$ years for SNO), we 
have computed the
sampling errors on the first two moments as well as the correlation of
the  errors. The iso-sigma ellipses for the
 six detectors we consider here are centered around the undistorted
zenith-angle exposure function for which, by definition,
 $\Delta\alpha = \Delta\sigma^2 =
0$. Figure~\ref{isosigma} 
shows for each detector the predicted shifts of the
first two moments in the SMA, LMA, and LOW solutions.
The horizontal and vertical error-bars denote
the spread in predicted values of the shifts in the first two moments,
which are obtained by varying $\Delta m^2$ and $\sin^2 2\theta$ within
the 95\% C.L. allowed (see Fig.~\ref{allowed}) by the four pioneering
solar neutrino experiments.

For Super-Kamiokande (SNO), the current best-fit parameters, 
$\Delta m^2$ and $\sin^22\theta$, predict a 
$5\sigma$ ($6.5\sigma$) effect for the SMA solution and $13\sigma$
($25\sigma$) effect for the LMA solution. 
Note that SNO is expected to require twice as much time to collect the
same number of events as Super-Kamiokande. In the same amount of
observing time, SNO and Super-Kamiokande are approximately equivalent
 for the SMA and Super-Kamiokande is significantly more efficient for the
LMA.

The current best-estimate MSW solutions predict
statistically significant deviations from the undistorted zenith-angle
moments for the Super-Kamiokande, SNO, and ICARUS experiments (which
are sensitive to the SMA and 
LMA solutions), but these experiments with the higher energy neutrinos
are not sensitive to the deviations predicted by the LOW solution. 
However, Figure~\ref{isosigma} shows that the 
BOREXINO and HERON/HELLAZ experiments are very sensitive to
the LOW solution.

\subsection{The Shape of the $^8$B Neutrino Energy Spectrum}
\label{shape}

The shape of the 
energy spectrum of neutrinos created by a specific continuum 
$\beta-$decay reaction is the same,
to an accuracy of order 1 part in $10^5$, for neutrinos that are
produced in the center of the sun and for neutrinos that are produced 
in a terrestrial laboratory, provided only that the minimal
standard electroweak theory is correct.\cite{Bahcall91} 
The physical reason for this result is
that the thermal velocities of ions in the solar interior are small 
compared to the velocity of light, $v^2/c^2 \sim 10^{-6}$.  First 
order corrections in $v/c$ vanish because the motions of the thermal 
ions are random. In fact, the largest correction ($\sim 10^{-5}$) 
to the shape of the
energy spectrum arises from the general relativistic redshift.\cite{Bahcall91} 

Given this result, it follows that a measurement of the shape of the 
$^8$B neutrino energy spectrum is direct test of minimal standard
electroweak theory. For small distortions, most of the available
information is contained in the value of the average electron recoil
energy, $<T_e>$ (see Appendix A of Bahcall and Lisi).\cite{snoJE}
If the distortion is large, it will show up clearly in any
characterization, including the average recoil energy.

\begin{figure}[tb]
\centerline{\psfig{figure=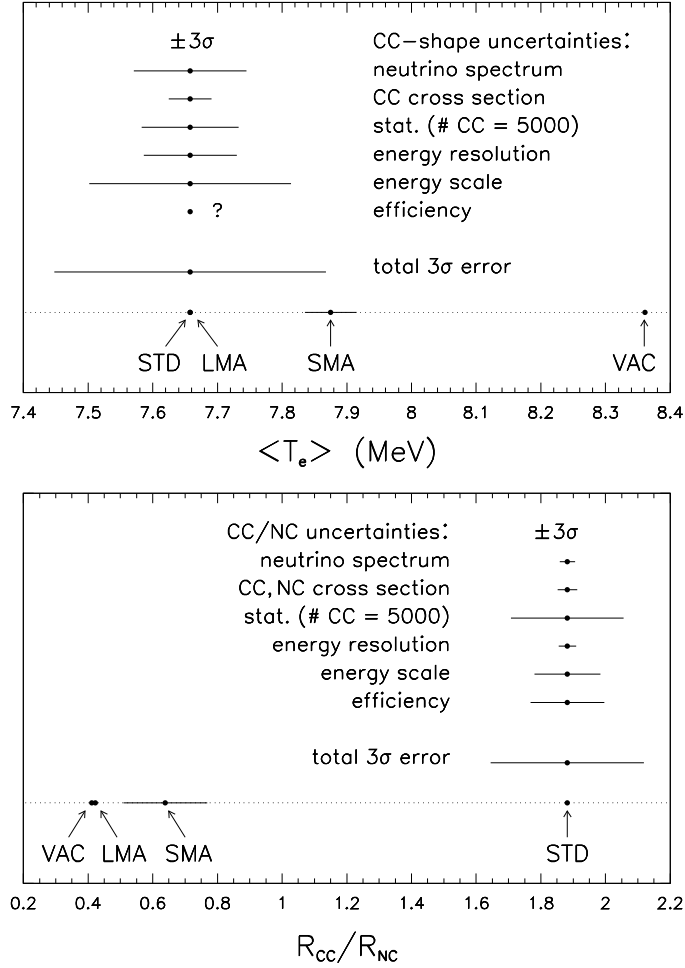,width=3.5in}}
\caption[]{\baselineskip=10pt\footnotesize Values 
of the characteristic CC-shape variable, the
average electron recoil energy $\langle T_e\rangle$ and of the CC/NC ratio, 
 $R_{\rm CC}/R_{\rm NC}$, together with  $3\sigma$
error bars. Here CC refers to $\nu_e$ absorption by deuterium with an
electron being produced, which occurs via the charged current.  The
neutral current, NC, disintegration of the deuteron occurs with an
equal cross section for all neutrino flavors ($\nu_e$, $\nu_\mu$, and
$\nu_\tau$). Uncertainties due to the backgrounds are
neglected.  This is Figure 7
of the paper on SNO by Bahcall and Lisi.\cite{snoJE}
\protect\label{snoT}}
\end{figure}

For SNO, Figure~\ref{snoT}a 
shows the  predictions for $\langle 
T_e\rangle$ that follow from the best-estimate 
 small angle (SMA) and large angle
(LMA) MSW solutions, as well as the vacuum (vac) oscillation solution.
The figure also shows 
the separate and combined $3\sigma$ errors expected from different
sources; the efficiency error (labeled by a question mark) should
be negligible if SNO works as expected.

\begin{figure}[t]
\centerline{\psfig{figure=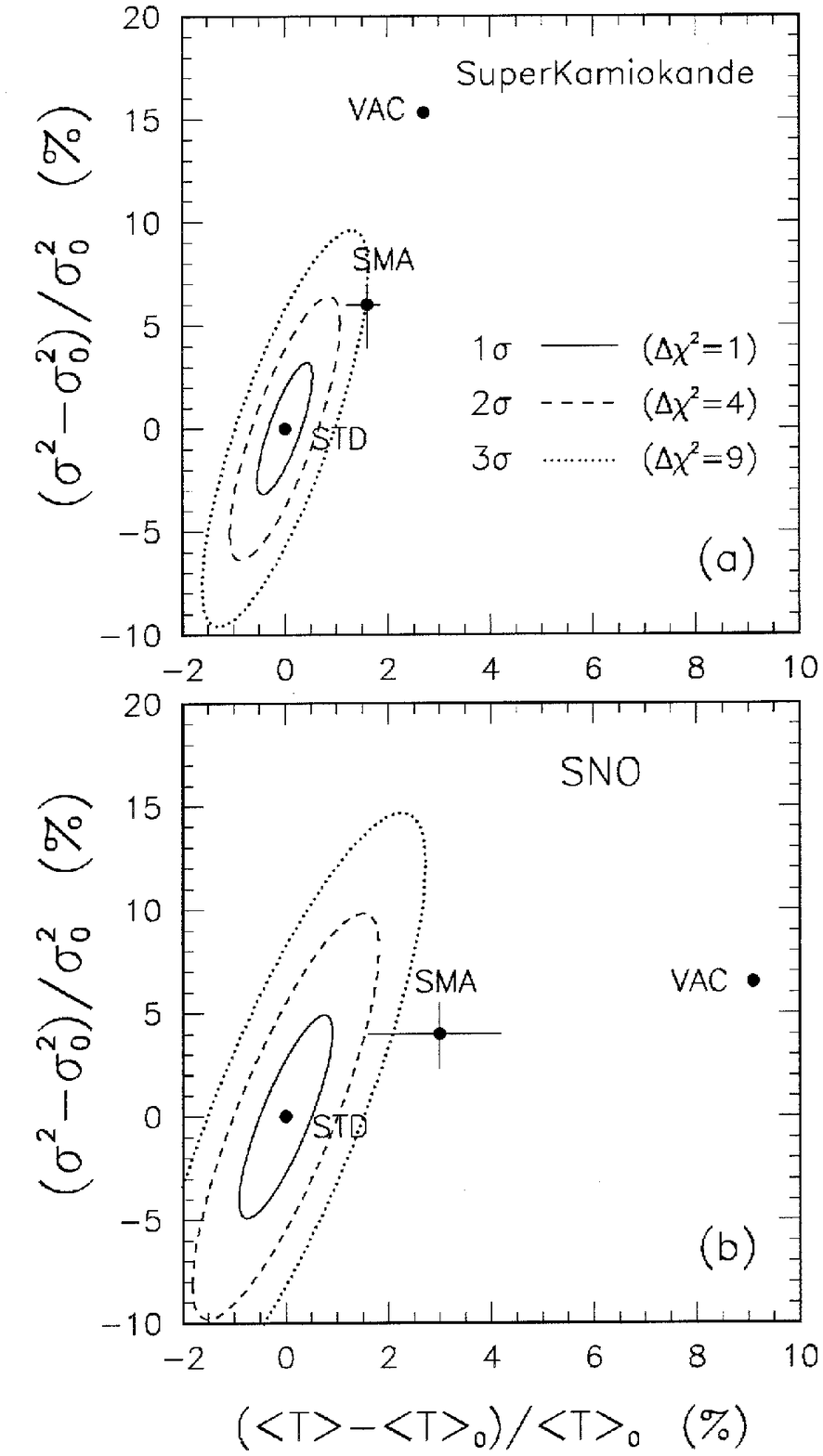,width=2.5in}}
\caption[]{\baselineskip=10pt\footnotesize 
Spectral energy distortion. Iso-sigma contours in the plane of the fractional deviations 
of the first two spectral moments. 
(a) Super-Kamiokande experiment. 
(b) SNO experiment. 
The SMA solution can be distinguished at 
$\protect\gtorder 3\sigma$ from the standard (STD) case
by both experiments. The crosses allow for variations of 
the SMA solution within the region  favored at $95\%$ C.L.\ 
by the current experiments. See the text for details. The figure is
Figure 4 from the paper of Bahcall, Krastev, and Lisi.\cite{BKL97}
\protect\label{isoSuperKSno}}
\end{figure}

Figure~\ref{isoSuperKSno} shows contours of equal standard deviations 
($n$-sigma ellipses)
in the plane of the $\langle T \rangle$ and $\sigma^2$ deviations 
of the spectrum that were 
computed for different neutrino scenarios. The contours are centered
around the standard expectations (STD). Also shown are the representative 
best-fit points VAC and SMA.  The point LMA is very close to STD and is 
not shown.

	The cross centered at the SMA best-fit point indicates the solution 
space allowed at  $95\%$ C.L. by the pioneering 
solar neutrino experiments.
The deviations 
in $\langle T \rangle$  and $\sigma^2$ for the SMA solution are confined 
to a  relatively small range. For vacuum oscillations, the range of 
deviations spanned by the whole region currently 
allowed at 95\% C.L.\ by present 
data  is  much larger and is not indicated 
in Fig.~\ref{isoSuperKSno}. 
The statistical significance of the separation between the SMA 
and STD points  in Fig.~\ref{isoSuperKSno}
 is dominated by the fractional shift in 
$\langle T \rangle$ for both Super-Kamiokande and SNO. 
This is not surprising, 
since the SMA neutrino survival probability increases almost linearly 
with energy for $E_\nu > 5$ MeV; this increase  induces deformations 
of the electron recoil spectrum that are nearly linear in $T$ and are well 
represented by a shift in $\langle T \rangle$ .

	The best-fit small mixing angle solution is separated by about
$3\sigma$ or more from the standard solution for both Super-Kamiokande
and SNO (see Fig.~\ref{isoSuperKSno}). 
The discriminatory power of the  two experiments 
appears to be comparable for the SMA solution. The  estimated total 
fractional errors of $\langle T \rangle$ and $\sigma^2$ in Super-Kamiokande 
are about a factor of two smaller than in SNO.
However, the purely charged current (CC) interaction in SNO 
($\nu_e$ absorption in deuterium)
is a more sensitive probe of neutrino oscillations than a linear 
combination of charged current  and neutral current (NC) 
interactions, as observed
in Super-Kamiokande.
In practice, the separation of charged current events and neutral current 
events (neutrino disintegration of the deuteron) in SNO will
be affected by experimental uncertainties. We  ignored
misidentifications in our simulations.

How do the above results depend upon the energy threshold?
The threshold is one of the most important 
quantities which experimentalists can hope to
improve in order to increase the sensitivity of their detectors to 
distortion of the energy spectrum.
We have determined by detailed calculations that 
the statistical 
significance of the SMA deviations in Figs.~4(a) and 4(b) decreases by 
about $0.6\sigma$ per 1 MeV increase in the  energy threshold 
$T_{\rm min}$.  These results are valid for both the SNO and the
Super-Kamiokande detectors and include calculations for thresholds of
$5$, $6$, and $7$ MeV.

\subsection{The CC to NC Ratio}
\label{cctoncratio}

The bottom line for nearly all of the particle physics descriptions of
what is happening in solar neutrino experiments is that a significant
fraction of the $\nu_e$'s that are created in the interior of the sun
are converted into $\nu_\mu$'s or $\nu_\tau$'s, either in the sun or
on the way to the earth from the sun.  The most direct test of this
deviation from minimum standard electroweak theory is to measure the 
ratio of the flux of $\nu_e$'s (via a charged current, CC, 
interaction) to the 
flux of neutrinos of all types ($\nu_e + \nu_\mu + \nu_\tau$,
determined 
by a neutral current, NC, interaction).
The SNO collaboration is completing the construction of a 1000 ton 
heavy water  
detector in the Creighton Mine (Walden, Canada).\cite{SNOw}  The detector
will measure the rates of the charged (CC) and neutral (NC) current reactions
induced by solar neutrinos in deuterium:
\begin{equation}
\nu_e + d \rightarrow p+p+e^-\quad({\rm CC\ absorption})\quad,
\label{reactionCC}
\end{equation}
\begin{equation}
\nu_x + d \rightarrow p+n+\nu_x\quad({\rm NC\ dissociation})\quad,
\label{reactionNC}
\end{equation}
including the determination of the electron recoil energy in 
Equation~(\ref{reactionCC}). Only the more energetic $^8$B solar neutrinos
are expected to be detected since the expected
SNO threshold  for CC events is an electron kinetic energy
of about 5 MeV and the physical threshold for NC dissociation is the 
binding energy of the deuteron, $E_b= 2.225$ MeV.

\begin{figure}[tb]
\vglue1in
\centerline{\psfig{figure=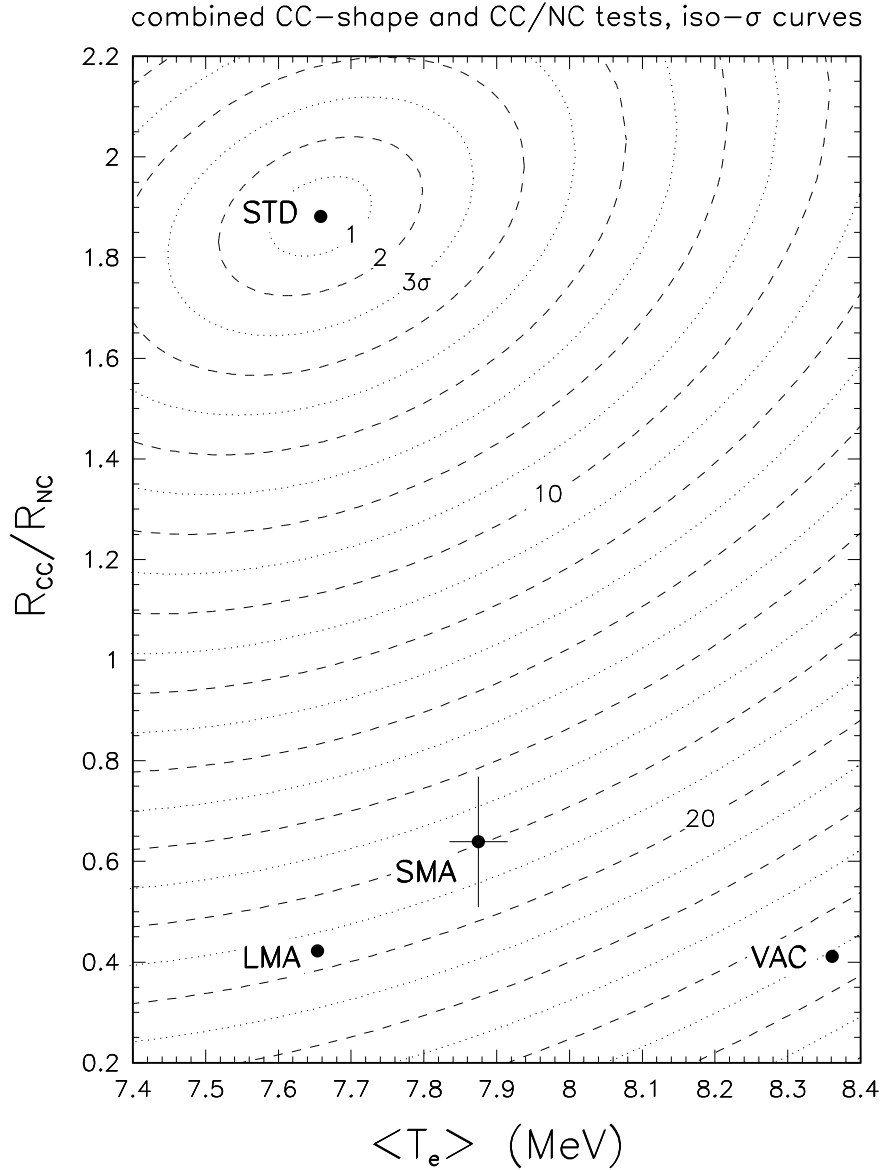,width=4.5in}}
\caption[]{\baselineskip=10pt\footnotesize Iso-sigma contours $(\sigma=\protect\sqrt{\chi^2})$
for the combined CC-shape and CC/NC test, for the 
representative oscillation cases 
 discussed in the text. 
Uncertainties due to the backgrounds are neglected. 
For values of the iso-sigma distance 
${\cal  N}(\sigma) \gg 3$, 
the number of standard deviations is only a formal 
characterization; the tail of the
probability distribution is not expected to be Gaussian 
for very large values of ${\cal
N}(\sigma)$. This is Figure~8 of Bahcall and Lisi.\cite{snoJE}
\protect\label{snoavTccnc}}
\end{figure}

Figures~\ref{snoT}a and \ref{snoT}b  
 show the standard predictions for the average recoil energy, $\langle 
T_e\rangle$ (upper panel), and for the ratio, $R_{\rm CC}/R_{\rm NC}$ (lower 
panel), of CC $(\nu_e)$ to NC (all flavors) 
together with the separate and combined $3\sigma$ errors. The 
values of $\langle T_e\rangle$ and  $R_{\rm CC}/R_{\rm NC}$ 
for the different oscillation channels are also displayed.

Figure~\ref{snoavTccnc}   shows the results of the combined tests 
(correlations included) in terms of iso-sigma contours in the plane 
$(\langle T_e\rangle,\,R_{\rm CC}/R_{\rm NC})$, where ${\cal 
N}(\sigma)=\sqrt{\chi^2}$.   The three oscillation scenarios can be well 
separated from the standard case, but the vertical separation 
($R_{\rm CC}/R_{\rm NC}$) is larger and dominating\ with respect to  
the horizontal separation ($\langle T_e\rangle$).

The error bars on the SMA point in Figs.~\ref{snoT}--\ref{snoavTccnc} 
  represent the  range of values allowed at 
$95\%$ C.L.\ by a fit of the oscillation predictions to the four 
pioneering solar neutrino experiments;\cite{BK97,snoJE,BKL97} 
the error bars  are intended to 
indicate  the effect of the likely range of 
the allowed oscillation parameters.

\subsection{The Seasonal Dependence of the Neutrino Fluxes}
\label{seasonal}

For vacuum neutrino oscillations, the survival probability of $\nu_e$
at a distance $L$ from the sun is given by\cite{Frautschi69}

\begin{equation}
P(E) = 1 - \sin^2 2\theta \left({1.27\Delta m^2{\rm (eV)}L(m)\over E {\rm(MeV)}}\right),
\label{vacuumsurvival}
\end{equation}
where $E$ is the neutrino energy, $\Delta m^2$ is the neutrino squared
mass-difference, and $\theta$ is the vacuum mixing angle.  The
ellipticity of the Earth's orbit implies a striking signature of the
oscillation phenomenon, namely, a dependence of the observed rate upon
the instantaneous earth-sun distance, $L$ (in addition to the trivial
geometric factor of $L^{-2}$). To a high  accuracy, $L(t) =
L_0\bigl( 1 - \epsilon \cos{{2 \pi t} \over T} \bigr)$. where $L_0$ is
$1$ AU, $T = 1$ yr, and $\epsilon = 0.0167$.  The periodic dependence
of the distance $L(t)$ upon time of the year implies a seasonal
variation of the neutrino event rates.\cite{pontecorvo,Fogli97,KP97}  
This variation is especially
noticeable for neutrino masses in the range of $10^{-10} {\rm eV}^2$,
which is consistent with some fraction (see Figure~2 of Bahcall and
Krastev\cite{BaKr})
of the vacuum neutrino solutions that
describe successfully the results of the pioneering solar neutrino
experiments. 

Among the second generation of experiments, the situation is most
favorable for the BOREXINO experiment, since the events in this
experiment are expected to be dominated by the $^7$Be (practically
monoenergetic) neutrino line. Large effects can be anticipated for
favorable cases for BOREXINO, but the effects will be reduced in the
Super-Kamiokande and SNO experiments because the rates in these 
experiments average  over neutrino energies. Fogli, Lisi, and
Montanino\cite{Fogli97} 
 propose a Fourier analysis of the neutrino signals for these
experiments and show that with $10^4$ events and no appreciable 
backgrounds (a very
optimistic assumption) there are currently-allowed vacuum neutrino
solutions that would produce a $3\sigma$ effect in the
Super-Kamiokande experiment and a $7\sigma$ effect in SNO.

\section{Summary of Second Lecture}
\label{summarytwo}

This is an incredibly exciting time to be doing solar neutrino
research. There is a widespread feeling among people working in the
field that we may be on the verge of making important discoveries
about how neutrinos behave.  

The greatest  concern I have is that there are too 
few experiments. Looking
back at the history of science, we see that it is necessary to have
redundant experiments in order to test whether or not unrecognized
systematic errors have crept into even the most careful measurements.

Only one experiment is planned that will measure a
neutral current reaction (the SNO measurement of deuteron
disintegration, see Equation~\ref{reactionNC}).  The neutral current to charged
current ratio of fluxes 
determines most directly what we need to know in order to decide if
new physics is occurring: the ratio
of the total number of neutrinos to the number of $\nu_e$'s.
Similarly, in order to test the
astronomical predictions for the number of neutrinos created in the
solar interior.
we must know the total number of neutrinos that reach the earth
in any flavor state. 

Of the funded experiments, only BOREXINO has the planned sensitivity
to detect the important $^7$Be neutrino line at $0.86$ MeV.  The
$^7$Be  line
is crucial for both the astronomical and the  physical interpretations of the
combined set of solar neutrino experiments (see for example the
discussion in my reviews\cite{bahcall96,bahcall97}).

There are currently no funded projects for measuring individual events
from the 
$pp$ neutrinos, although both HELLAX and HERON seem very promising.
The  low-energy $pp$ neutrinos constitute more than $90$\%
of the total solar neutrino flux in standard models. 
The radiochemical experiments,
GALLEX, SAGE, and GNO,  give us fundamental 
 upper limits to the $pp$ flux
at earth, but to exploit fully either the solar or 
the physics information encoded in the
$pp$ neutrino flux we need measurements which determine the energy
associated with each observed neutrino event.

We need more experiments, especially experiments sensitive to neutrinos
with energies below $1$
MeV and experiments sensitive to neutral currents.

\section*{Acknowledgments}
This research is supported in part by NSF
grant number PHY95-13835.

\end{document}